\documentclass[a4paper]{jpconf}

\usepackage{amsfonts,amsmath,amssymb,bm,hyperref}

\begin{document}
\title{Neutrino flavor oscillations in background matter}

\author{Maxim Dvornikov}

\address{Department of Physics, P.O. Box 35, FIN-40014, University of
Jyv\"{a}skyl\"{a}, Finland; \\
IZMIRAN, 142190, Troitsk, Moscow region, Russia}

\ead{dvmaxim@cc.jyu.fi; maxdvo@izmiran.ru}

\begin{abstract}
We study the evolution of two mixed Dirac neutrinos in an external axial-vector 
field. The dynamics of this system is described in the framework of 
relativistic wave equations approach on the basis of the known solutions of the 
Dirac equation in an external field with a given initial condition. The general 
solution of the initial condition problem, exactly accounting for the external 
field, is obtained. Then we consider special form of the external field which 
corresponds to the standard model neutrino interactions with the background 
matter. In this limit we derive the transition probability and compare it with 
the previously known results.
\end{abstract}

The problem of neutrino oscillations in various external fields attracts 
considerable attention mainly because the amplification of neutrino 
oscillations in background matter (MSW effect, see 
Refs.~\cite{Wol78,MikSmi85eng}) is the most plausible explanation of the solar 
neutrino deficit. We discussed neutrino flavor and spin-flavor oscillations in 
our recent works~\cite{Dvo05,Dvo06EPJC,DvoMaa07,Dvo07} in frames of the 
relativistic wave equations approach (classical field theory) which exactly 
accounts for external fields. 
The present paper continues our previous efforts.      

Let us discuss the evolution of two mixed flavor neutrinos $\nu_\lambda$, 
$\lambda=\alpha,\beta$, interacting with the external field axial-vector field 
$f_{\lambda\lambda'}^{\mu}$. We assume that the matrix 
$(f_{\lambda\lambda'}^{\mu})$ is generally not diagonal in the indexes 
$\lambda,\lambda'$. Note that the similar external fields were recently 
discussed in Ref.~\cite{KopLinOta07} (see also references therein). The 
Lagrangian describing the dynamics of the considered system is
\begin{equation}\label{Lagrnu}
  \mathcal{L}(\nu_\alpha,\nu_\beta)=
  \sum_{\lambda=\alpha,\beta}
  \bar{\nu}_\lambda \mathrm{i}\gamma^\mu\partial_\mu \nu_\lambda-
  \sum_{\lambda,\lambda'=\alpha,\beta}
  \left[
    m_{\lambda\lambda'} \bar{\nu}_\lambda \nu_{\lambda'}+	 
	f_{\lambda\lambda'}^{\mu}
	\bar{\nu}_\lambda \gamma_\mu^\mathrm{L} \nu_{\lambda'}
  \right],
\end{equation}
where $(m_{\lambda\lambda'})$ is the mass matrix of flavor neutrinos and 
$\gamma_\mu^\mathrm{L}=\gamma_\mu(1+\gamma^5)/2$. Note that the matrices 
$(m_{\lambda\lambda'})$ and $(f_{\lambda\lambda'}^{\mu})$ are generally 
independent. We supply the Lagrangian~\eqref{Lagrnu} with the initial
conditions, $\nu_{\alpha}(\mathbf{r},0)=0$ and
$\nu_{\beta}(\mathbf{r},0)=\xi(\mathbf{r})$,
where $\xi(\mathbf{r})$ is a given function. This situation is implemented, 
e.g., if $\nu_\alpha \equiv \nu_\mu$, $\nu_{\beta} \equiv \nu_e$ and we study 
oscillations of solar neutrinos, when only electron neutrinos are presented 
initially. 

To examine the evolution of the system~\eqref{Lagrnu} we introduce the mass 
eigenstates $\psi_a(\mathbf{r},t)$, $a=1,2$, by means of the matrix 
transformation, $\nu_{\lambda}=\sum U_{\lambda a}\psi_a$. In case of two flavor 
neutrinos the matrix $(U_{\lambda a})$ is the rotation matrix in the flavor 
basis, $U=\exp(-\mathrm{i} \sigma_2 \theta)$, where $\theta$ is the vacuum 
mixing angle. The effective Lagrangian for the mass eigenstates is obtained 
from Eq.~\eqref{Lagrnu},
\begin{equation}\label{Lagrpsi}
  \mathcal{L}(\psi_1,\psi_2)=
  \sum_{a=1,2}\bar{\psi}_a(\mathrm{i}\gamma^\mu \partial_\mu-m_a)\psi_a-
  \sum_{a,b=1,2}
  g_{ab}^{\mu} \bar{\psi}_a \gamma_\mu^\mathrm{L} \psi_{b},
\end{equation}
where $m_a$ the mass is fermion $\psi_a$ and 
$(g_{ab}^{\mu})=U^{-1}(f_{\lambda\lambda'}^{\mu})U$ is the external 
axial-vector field expressed in the mass eigenstates basis. 

The Dirac equations which result from Eq.~\eqref{Lagrpsi} have the form,
\begin{equation}\label{psiHamform}
  \mathrm{i}\dot{\psi}_a=\mathcal{H}_a\psi_a+V\psi_b,
  \quad
  a,b=1,2,
  \quad
  a \neq b,
\end{equation}
where $\mathcal{H}_a=-\mathrm{i}\bm{\alpha}\mathbf{\nabla}+\beta m_a + 
\beta\gamma_\mu^\mathrm{L} g_a^\mu$, $V=\beta\gamma_\mu^\mathrm{L} g^\mu$, 
$g_a^\mu = g_{aa}^\mu$ and $g^\mu=g_{12}^\mu$.
The general solution to Eq.~\eqref{psiHamform} can be presented in the 
following way (see Refs.~\cite{Dvo06EPJC,DvoMaa07,Dvo07}):
\begin{equation}\label{gensol}
  \psi_{a}(\mathbf{r},t)=
  \int \frac{\mathrm{d}^3\mathbf{p}}{(2\pi)^{3/2}}
  e^{\mathrm{i}\mathbf{p}\mathbf{r}}\sum_{\zeta=\pm 1}
  \left[
    a_a^{(\zeta)}(t)u_a^{(\zeta)}\exp{(-\mathrm{i}E_a^{(\zeta)} t)}+
	b_a^{(\zeta)}(t)v_a^{(\zeta)}\exp{(+\mathrm{i}E_a^{(\zeta)} t)}
  \right],
\end{equation}
where $a_a^{(\zeta)}$ and $b_a^{(\zeta)}$ are the undetermined 
\emph{non-operator} coefficients, which are generally time depedent. The energy 
spectrum $E_a^{(\zeta)}$ in case of non-moving and unpolarized mater, which 
corresponds to $\mathbf{g}_a=\mathbf{g}=0$, was found in 
Refs.~\cite{StuTer05,Lob05},
\begin{equation}\label{energy}
  E_a^{(\zeta)}=
  \sqrt{\mathbf{p}^2
  \left(
    1-\zeta\frac{g_a}{2|\mathbf{p}|}
  \right)^2+m_a^2}+
  \frac{g_a}{2}.
\end{equation}
In Eq.~\eqref{energy} we introduce the quantity $g_a \equiv g_a^{0}$. The basis 
spinors $u_a^{(\zeta)}$ and $v_a^{(\zeta)}$ in Eq.~\eqref{gensol} are the 
eigenvectors of the helicity operator $(\bf{\Sigma}\mathbf{p})/|\mathbf{p}|$, 
with the eigenvalues $\zeta = \pm 1$, and can be also found in the explicit 
form in Refs.~\cite{StuTer05,Lob05}.

Let us assume the initial wave function in the form, 
$\xi(\mathbf{r})=e^{\mathrm{i}\mathbf{k}\mathbf{r}}\xi_0$, where 
$\mathbf{k}=(0, 0, k)$. If we study relativistic neutrinos with $k\gg m_{1,2}$, 
the normalized basis spinors in Eq.~\eqref{gensol} are
\begin{equation}\label{spinors}
  u^{+{}}=
  \frac{1}{\sqrt{2}}
  \begin{pmatrix}
     1 \\
     0 \\
	 1 \\
	 0 \ 
  \end{pmatrix},
  \quad
  u^{-{}}=
  \frac{1}{\sqrt{2}}
  \begin{pmatrix}
     0 \\
     -1 \\
	 0 \\
	 1 \ 
  \end{pmatrix},
  \quad
  v^{+{}}=
  \frac{1}{\sqrt{2}}
  \begin{pmatrix}
     1 \\
     0 \\
	 -1 \\
	 0 \ 
  \end{pmatrix},
  \quad
  v^{-{}}=
  \frac{1}{\sqrt{2}}
  \begin{pmatrix}
     0 \\
     1 \\
	 0 \\
	 1 \ 
  \end{pmatrix}.
\end{equation}
One can take that $\xi_0=u^{-{}}$. It is easy to check that 
$(1/2)(1-\Sigma_3)\xi_0=\xi_0$. Therefore $\xi(\mathbf{r})$ describes a 
neutrino propagating along the $z$-axis, with the spin directed opposite to the 
particle momentum. Note that the subscript $a$ is omitted in 
Eq.~\eqref{spinors} since we neglect small terms $\sim m_a/k \ll 1$.

With help of the obvious identities, 
$\langle u^{-{}} | V | u^{-} \rangle = \langle v^{+{}} | V | v^{+{}} \rangle = 
g^0$ (all other matrix elements of the potential $V$ vanish), which result from 
Eq.~\eqref{spinors}, as well as using the technique developed in 
Refs.~\cite{DvoMaa07,Dvo07} we get the ordinary differential equations for the 
functions $a^{-{}}_a(t)$,  

\begin{equation}\label{aeq}
  \mathrm{i}\dot{a}_a^{-{}}= 
  a_b^{-{}} g
  \exp{[\mathrm{i}(E_a^{-{}}-E_b^{-{}}) t]},
  \quad
  a,b=1,2,
  \quad
  a \neq b,
\end{equation}
where $g \equiv g^{0}$. Note that the equations for the functions 
$b^{+{}}_a(t)$ are obtained analogously. The solution to Eq.~\eqref{aeq} can be 
expressed in the form (see, e.g., Refs.~\cite{DvoMaa07,Dvo07}) 

\begin{alignat}{2}\label{aeqsol}
  a_1^{-{}}(t)= & 
  F a_1^{-{}}(0)+G a_2^{-{}}(0), &
  \qquad
  F = &
  \left[
    \cos\Omega t-
    \mathrm{i}\frac{\omega}{2\Omega}\sin\Omega t
  \right]
  \exp{(\mathrm{i}\omega t/2)},
  \notag
  \\
  a_2^{-{}}(t)= & 
  F^{*{}}a_2^{-{}}(0)-G^{*{}}a_1^{-{}}(0), &
  G = &
  -\mathrm{i}\frac{g}{\Omega}\sin\Omega t
  \exp{(\mathrm{i}\omega t/2)},
\end{alignat}
where $\Omega=\sqrt{g^2+(\omega/2)^2}$ and $\omega=E_1^{-{}}-E_2^{-{}}$.

Using the identity $\left(v^{+{}} \otimes v^{+{}\dag}\right)\xi_0=0$ [see 
Eq.~\eqref{spinors}] as well as Eqs.~\eqref{gensol} and~\eqref{aeqsol} we 
arrive to the wave function of the neutrino $\nu_\alpha$,
%
%
\begin{equation}\label{nualpha}
  \nu_\alpha(z,t) =
  -\mathrm{i}
  \exp{(-\mathrm{i}\bar{\mathcal{E}}t+\mathrm{i}kz)}
  \sin\Omega t
  \frac{[g\cos 2\theta+(\omega/2)\sin 2\theta]}{\Omega}\xi_0+
  \mathcal{O}
  \left(
    \frac{m_a}{k}
  \right),
\end{equation}
where $\bar{\mathcal{E}}=(E_1^{-{}}+E_2^{-{}})/2$. Note that 
Eq.~\eqref{nualpha} is the most general one which is valid for the external 
axial-vector fields $f_{\lambda\lambda'}^\mu$ of arbitrary strength. Let us 
however discuss one of the applications of the obtained result. We consider 
diagonal matrix 
$f_{\lambda\lambda'}^\mu=f_\lambda^{\mu}\delta_{\lambda\lambda'}$ which 
corresponds to the standard model neutrino interactions with the non-moving and 
unpolarized background matter, i.e. we take that $\mathbf{f}_\lambda=0$. Note 
that the values of $f_\lambda^0$ for various channels of neutrino oscillations 
can be found in Ref.~\cite{DvoStu02JHEP} In this case we have $g = -\sin 
2\theta \Delta f^0$, where $\Delta f^0 = (f_\alpha^0-f_\beta^0)/2$. In the low 
density matter limit, $g_a \ll k$, one reads $\omega/2 \approx \Phi(k)+\cos 
2\theta \Delta f^0$ [see Eq.~\eqref{energy}], where $\Phi(k)=\delta m^2/(4k)$ 
and $\delta m^2=m_1^2-m_2^2$.

The transition probability can be calculated on the basis of 
Eq.~\eqref{nualpha} as
\begin{equation}\label{Ptr}
  P_{\nu_\beta\to\nu_\alpha}(t) =
  |\nu_\alpha(z,t)|^2 \approx
  A\sin^2
  \left(
    \frac{\pi}{L}t
  \right),
\end{equation}
where
\begin{equation}\label{parameters}
  A=
  \frac{\Phi^2(k)\sin^2(2\theta)}
  {[\Phi(k)\cos2\theta+\Delta f^0]^2+
  \Phi^2(k)\sin^2(2\theta)},
  \
  \frac{\pi}{L}=
  \sqrt{[\Phi(k)\cos2\theta+\Delta f^0]^2+
  \Phi^2(k)\sin^2(2\theta)}.
\end{equation}
We can observe that Eqs.~\eqref{Ptr} and~\eqref{parameters} reproduce the 
famous formula for the neutrino oscillations probability in the background 
matter (see Refs.~\cite{Wol78,MikSmi85eng}).

In conclusion we mention that neutrino flavor oscillations in background matter 
with non-standard interactions have been studied in frames of the relativistic 
wave equations approach (classical field theory). Neutrino interactions with 
matter are known to be equivalent to the presence of an external axial-vector 
field. In the limit of ultrarelativistic neutrinos interacting with non-moving 
and unpolarized matter we have received the most general expression for the 
neutrino wave function [Eq.~\eqref{nualpha}], which exactly takes into account 
the external axial-vector field. Using this result and considering standard 
model neutrino matter interactions we could reproduce the transition 
probability [Eqs.~\eqref{Ptr} and~\eqref{parameters}] accounting for the 
resonance enhancement of neutrino oscillations -- the MSW effect. Note that 
with help of the method used in the present work we have improved the results 
of our recent paper~\cite{Dvo06EPJC}. In contrast to that work, where the case 
of only low density matter was discussed, now we could obtain the transition 
probability valid for arbitrary matter density. 

\section*{Acknowledgments}
The work has been supported by the Academy of Finland under the
contract No.~108875. The author is thankful to the Russian Science Support 
Foundation for a grant and the organizers of the EPS HEP 2007 conference for 
the invitation.

\section*{References}

\bibliography{generaleng}

\end{document}